\documentclass[cmp,draft,numbook]{svjour}
\usepackage{amsmath,amssymb,verbatim}

\spnewtheorem*{cor}{Corollary}{\bf}{\it}
\spnewtheorem*{lem}{Lemma}{\bf}{\it}
\spnewtheorem*{pro}{Proposition}{\bf}{\it}
\spnewtheorem{teo}{Theorem}{\bf}{\it}

\newcommand\dd{\partial}

\newcommand\de{\delta}
\newcommand\FF{\mathbb F}
\renewcommand\ge{\geqslant}
\newcommand\La{\Lambda}
\newcommand\la{\lambda}

\newcommand\lcd{,\ldots,}
\renewcommand\le{\leqslant}

\newcommand\QQ{\mathbb Q}
\renewcommand\phi{\varphi}

\newcommand\ts{\hspace{0.75pt}}

\newcommand\D{\mathcal{D}}
\newcommand\Lc{\mathcal{L}}
\newcommand\Q{\mathcal{Q}}
\newcommand\Qa{\Q^{\ts\ast}}
\newcommand\U{\mathcal{U}}
\newcommand\Ua{\U^{\ts\ast}}

\begin{document}

\title{Lax operator for Macdonald symmetric functions}
\titlerunning{Lax operator}

\author{M.\,L.\,Nazarov and E.\,K.\,Sklyanin}
\authorrunning{Nazarov and Sklyanin}

\institute{Department of Mathematics, University of York, 
York YO10 5DD, United Kingdom}

\date{}%{31 October 2014}

\maketitle

%------------------------------------------------------------------------------

\thispagestyle{empty} % NO FIRST PAGE NUMBER

\begin{abstract}
Using the Lax operator formalism, 
we construct a family of pairwise commuting operators such that 
the Macdonald symmetric functions of infinitely many variables 
$x_1,x_2,\,\ldots$ and of two parameters $q,t$
are their eigenfunctions.
We express our operators in terms of
the Hall\ts-Littlewood symmetric functions
of the variables $x_1,x_2,\,\ldots\ $
and of the parameter $t$
corresponding to the partitions with one part only.
Our expression is based on the notion of Baker\ts-Akhiezer function.
\end{abstract}


%MSC2010: Primary 05E05; Secondary 17B69, 05E10

\newpage%%%%%%%%%%%%%%%%%%%%%%%%%%%%%%%%%%%%%%%%%%%%%%%%%%%%%%%%%%%%%%%%%%%%%%%

%==============================================================================

\section*{Introduction}

Over the last two decades
the Macdonald polynomials \cite{M}
have beeen
the subject of 
much attention
in combinatorics and representation theory.
These polynomials are symmetric in $N$ variables
$x_1\lcd x_N$ and also depend on two parameters %usually
denoted by $q$ and $t\,$. They are labelled by partitions of $0,1,2,\ldots$
with no more than $N$ %non-zero 
parts. Up to normalization, 
they can be defined
as eigenvectors of certain family of commuting
linear operators
acting on the space $\La_N$ of all symmetric polynomials in the variables 
$x_1\lcd x_N$ with coefficients
from the field $\QQ(q,t)\,$.
These operators were introduced by Macdonald \cite{M}
as the coefficients of a certain operator valued polynomial of degree $N$
in a variable $u$ with the constant term~$1\ts$.
In particular, Macdonald observed that all %the 
eigenvalues of the operator coefficient
at $u$ are already free from multiplicities. 
Hence this coefficient alone can be used to define
the Macdonald polynomials.

It is quite common in combinatorics to extend various symmetric polynomials
to an infinite %countable 
set of variables. These extensions are called
\emph{symmetric functions}. The ring $\La$ 
of symmetric functions is defined as the inverse limit of~the sequence
$
%\La_{\ts0}\leftarrow
\La_1\leftarrow\La_2\leftarrow\ldots
$ 
in the category of graded algebras. The defining homomorphism
$\La_{N-1}\leftarrow\La_N$ here is just the substitution $x_N=0\,$.
In particular,
the Macdonald polynomials are extended to infinitely many
variables $x_1,x_2,\ldots$ by using their \textit{stability property}
\cite{M} and by passing to their limits as 
$N\to\infty\,$. The limits are the Macdonald symmetric functions.
They belong to the ring $\La$ and 
are labelled by partitions of 
$0,1,2,\ldots\,\,$.
They have been also studied very well.
In particular, the limit at $N\to\infty$
of the renormalized Macdonald operator coefficient at $u$
was considered in \cite{M}.
Other expressions for the same limit were given in
\cite{AMOS,CJ}.

Since the higher operator coefficients are not required to define
the Macdonald polynomials, the limits of these coefficients at $N\to\infty$
received due attention only later. From the geometric point of view
they were studied in \cite{SV2}, see also the works 
\cite{FHHSY,SV1,Y}.
Explicit expressions for the limits were given in \cite{AK,BGHT,S} 
and independently in \cite{NS2}. 
All these expressions involved Hall\ts-Littlewood 
symmetric functions \cite{M} in the variables
$x_1,x_2,\ldots\ $. These symmetric functions are
labelled by partitions of $0,1,2,\ldots$
but depend on the parameter $t$ only.
They also emerge in the calculus of 
\emph{vertex operators}~\cite{J1}
which underlies the results of \cite{AK,AMOS,CJ,FHHSY,S}. 

In the present article we construct a different family of
commuting operators on $\La$ 
such that the Macdonald symmetric 
functions are their eigenvectors. Unlike in \cite{AK,BGHT,NS2}
our operators are expressed in terms of 
the Hall\ts-Littlewood symmetric 
functions corresponding to the partitions with one part only. 
Our construction uses the Lax operator formalism, 
see Subsection 2.1 for details. 
Our Theorem~1 gives a relation between the new family
of commuting operators and the one we constructed in \cite{NS2}.
The proof is based on the notion of Baker\ts-Akhiezer~function 
corresponding to the Lax operator, see Subsection 2.2.
In our case this function is given by Theorem 2.
The proof of the latter theorem is given in Subsection~2.3.  

\enlargethispage{6pt}%%%%%%%%%%%%%%%%%%%%%%%%%%%%%%%%%%%%%%%%%%%%%%%%%%%%%%%%%%

To find the eigenvalues of our new
operators we still need the results of \cite{NS2}. 
It would be interesting to prove directly that the eigenvectors of these
operators are the Macdonald symmetric functions, see for instance \cite{J2}.
Also notice that by setting $q=t^{\ts \alpha}$ and tending $t\to 1$ 
one obtains the \emph{Jack symmetric functions} \cite{M}
%of the parameter $\alpha$
as limits of Macdonald symmetric functions.
Our new Lax operator can be regarded 
as a discretization of the operator we found in the limiting case 
\cite{NS1,NS3}. The latter operator has been in turn a quantized version 
of the Lax operator for the
classical Benjamin\ts-Ono equation \cite{B,O}. 
Our new Lax operator is
a quantized version of the one for the 
classical Benjamin\ts-Ono equation with discrete Laplacian
\cite{ST1,ST2}.

\newpage%%%%%%%%%%%%%%%%%%%%%%%%%%%%%%%%%%%%%%%%%%%%%%%%%%%%%%%%%%%%%%%%%%%%%%%

In this article we generally keep to
the notation of the book \cite{M} %of Macdonald 
for symmetric functions. When using results from \cite{M}
we simply indicate their numbers within the book.
For example, the statement (2.15) from Chapter III of the book
will be referred to as [III.2.15] assuming it is from~\cite{M}. 

%==============================================================================

\section{Symmetric functions}

%------------------------------------------------------------------------------

\subsection{Power sums}

Fix any field $\FF\,$. 
Denote by $\La$ the $\FF$-algebra of symmetric 
functions in infinitely many variables $x_1,x_2,\ldots\,\,$.
%Following \cite{M} we will introduce some standard bases of $\La\,$.
Let $\la=(\,\la_1,\la_2,\ldots\,)$ be any partition of $\,0,1,2,\ldots\,\,$. 
We will always assume that $\la_1\ge\la_2\ge\ldots\,\,$.
The number of non-zero parts is called the {\it length\/} of 
$\la$ and denoted by $\ell(\la)\,$. 
%We will also denote $|\la|=\la_1+\la_2+\ldots$ as usual.
Let $k_1,k_2,\ldots$ be
the multiplicities of the parts
$1,2,\ldots$ of $\la$ respectively. Then $k_1+k_2+\ldots=\ell(\la)\,$.

For $n=1,2,\ldots$ let $p_n\in\La$ be
the \textit{power sum symmetric function\/} of degree~$n\,$:
%By definition,
\begin{equation*}
\label{pon}
p_n=x_1^n+x_2^n+\ldots\,.
\end{equation*}
More generally, for any partition $\la$ put
$
%\label{pla}
p_\la=p_{\la_1}\ldots p_{\la_k}
$
where $k=\ell(\la)$. The elements $p_\la$
form a basis of $\La\,$. In other words,
the elements $p_1,p_2,\ldots$ are
free generators of the commutative algebra~$\La$ over $\FF\,$.

Define a bilinear form $\langle\ ,\,\rangle$ on %the vector space 
$\La$ by setting for any two partitions $\la$ and $\mu$
\begin{equation}
\label{schurprod}
\langle\,p_\la,p_\mu\ts\rangle=k_\la\ts\de_{\la\mu}
\quad\text{where}\quad
k_\la=1^{k_1}k_1!\,2^{k_2}k_2!\ts\,\ldots
\end{equation}
in the above notation. This form is obviously symmetric
and non-degenerate. We will indicate by the superscript ${}^\perp$
the operator conjugation relative to this form. In particular, by 
%the definition 
\eqref{schurprod}
for the operator conjugate to the multiplication in $\La$ by $p_n$
with $n\ge1$ %we have
\begin{equation}
\label{pperp}
p_n^{\ts\perp}=n\,\dd/\dd\,p_n\,.
\end{equation}

%------------------------------------------------------------------------------

\subsection{Elementary and complete symmetric functions}

For $n=1,2,\ldots$ let $e_n\in\La$ be
the \textit{elementary symmetric function\/} of degree $n\,$.
By definition,
\[
e_n\ =
\sum_{i_1<\ldots<i_k}
x_{i_1}\ldots x_{i_k}\,.
\]
We will also use a formal power series in the variable $z$
\[
E(z)=1+e_1\ts z+e_2\ts z^2+\ldots\,=\,\prod_{i\ge1}\,(1+x_i\ts z)\,.
\]
By taking logarithms of the left and right hand side
of the above display and~then exponentiating,
\begin{equation}
\label{euexp}
E(z)\,=\,\exp\,\Bigl(\ts-
\sum_{n\ge1}\,
\frac{p_n}n\ts (-z)^n
\ts\Bigr)\,.
\end{equation}

The \textit{complete symmetric functions\/} 
$h_1,h_2\ldots$ %of degrees $1,2,\ldots$ respectively 
can be determined by the relation
\begin{equation}
\label{eh}
E(-z)\ts H(z)=1
\end{equation}
where
$$
H(z)=1+h_1\ts z+h_2\ts z^2+\ldots\,.
$$
The degree of the element $h_n\in\La$ is $n\,$. Furthermore, by 
\eqref{euexp} we get an equality
\begin{equation}
\label{huexp}
H(z)\,=\,\exp\,\Bigl(\,\,
\sum_{n\ge1}\,
\frac{p_n}n\ts z^n
\ts\Bigr)\,.
\end{equation}
The elements $h_1,h_2,\ldots$ as well as the elements $e_1,e_2,\ldots$
are free generators of the commutative algebra~$\La$ over the field $\FF\,$.

%------------------------------------------------------------------------------

\subsection{Hall\ts-Littlewood functions}

Let $\FF$ be the field $\QQ(t)$ with $t$ a parameter.~Put
$$
Q(z)=E(-\ts t\ts z)\ts H(z)=1+Q_1\ts z+Q_2\ts z^2+\ldots\,.
$$
Note that then by \eqref{euexp} and \eqref{eh} we have
\begin{equation}
\label{qprod}
Q(z)\,=\,\prod_{i\ge1}\ts
\frac{\,1-t\,x_i\,z\,}{1-x_i\ts z}
\,=\,\exp\,\Bigl(\,\,
\sum_{n\ge1}
\frac{1-t^n\!}{n}\,\ts p_n\ts z^n
\ts\Bigr)\,.
\end{equation}
In this article we will employ the {\it Jing vertex operator}
\begin{equation}
\label{ju}
J(z)=Q(z)\,E^\perp(-z^{-1})
\,.
\end{equation}
This is a formal series in $z$ with coefficients acting on $\La$ as
linear operators. These operators do not commute, see 
\cite[Proposition 2.12]{J1} for commutation relations
between them.
%For any $n\in\ZZ$ let $J_n$ be the coefficient of 
%the series \eqref{ju} at $z^n$.
%Let us apply these operator
%coefficients to the elements $ Q_1,Q_2\ts,\,\ldots\in\La\,$.
Using another variable $w$ instead of $z$ in 
the equalities~\eqref{qprod}~we~get
\begin{gather}
\notag 
E^\perp(-z^{-1})\ts(\ts Q(w))=
\exp\,\Bigl(\ts-
\sum_{n\ge1}\,
z^{-n}\,\dd/\dd\ts p_n
\ts\Bigr)
(\ts Q(w))=
\\
Q(w)\ts
\exp\,\Bigl(\ts-
\sum_{n\ge1}
\frac{1-t^n\!}{n}\,z^{-n}\ts w^n
\ts\Bigr)
=Q(w)\,
\frac{z-w}{z-t\,w}
\label{jq}
\end{gather}
due to \eqref{pperp} and \eqref{euexp}. The fraction 
at the right hand side of the equalities \eqref{jq}
should be expanded as a power series in the ratio $w/z\,$. 
It follows by \eqref{eh} that
\begin{equation}
\label{hq}
H^\perp(z^{-1})\ts(\ts Q(w))=
Q(w)\,\frac{z-t\,w}{z-w}\ .
\end{equation}

Following \cite[Proposition 3.9]{J1}  we will use the relation
\begin{equation}
\label{qkz}
J(z)\ts(\ts Q(z_1)\ldots Q(z_k))\,=\,Q(z)\,Q(z_1)\ldots Q(z_k)\,\ts
\prod_{j=1}^k\,\frac{z-z_j}{z-t\,z_j}\ .
\end{equation}
To prove \eqref{qkz} note that due to [Ex.\,I.5.25]
the series $E^\perp(-z^{-1})$ with operator coefficients 
showing in the definition  \eqref{ju} is comultiplicative, so that
$$
E^\perp(-z^{-1})\ts(\ts Q(z_1)\ldots Q(z_k))=
E^\perp(-z^{-1})\ts(\ts Q(z_1))\ts\ldots\ts E^\perp(-z^{-1})\ts(\ts Q(z_k))\,.
$$
Hence the equality
\eqref{qkz} is obtained by using \eqref{jq} with $w=z_1\lcd z_k\,$.
  
%Due to \cite[Proposition 3.9]{J1} 
%can be determined recursively by setting
%\begin{equation}
%\label{qla}
%Q_\la=J_{\la_1}\ldots J_{\la_k}(1)\,.
%\end{equation}
%We consecutively apply 
%the operators $J_{\la_k}\lcd J_{\la_1}$
%to the identity element $1\in\La\,$.
%If $k=0$ so that the partition $\la$ is empty, 
%then $Q_\la=1$ by definition. If $\la$ has only one part $n\ge1\ts$,
%then $Q_\la=Q_n\ts$. This is because
%$e_{\ts1}^\perp(1)=e_{\ts2}^\perp(1)=\ldots=0\,$.
%This definition of $Q_\la\in\La$ is equivalent to the classical one
%[III.2.15]. To state the latter for $k\ge1$

Now let $\la$ be any partition with $\ell(\la)=k\ts$.
Recall that $\la_1\ge\ldots\ge\la_k$ by our assumption.
Introduce a rational function of the variables $z_1\lcd z_k$
\begin{equation}
\label{fuk}
F(z_1\lcd z_k)\,=\prod_{1\le i<j\le k}\frac{z_i-z_j}{\,z_i-t\,z_j}\ .
\end{equation}
Let us expand every factor with $i<j$
in the product \eqref{fuk}
as a power series in $z_j/z_i$ respectively.
By [III.2.15] the \emph{Hall\ts-Littlewood symmetric function\/}
$Q_\la\in\La$ is the coefficient at %the monomial
$$
z_1^{\ts\la_1}\dots z_k^{\ts\la_k}
$$
in the formal series 
$$
Q(z_1)\ldots Q(z_k)\,
F(z_1\lcd z_k)\,.
$$
If the partition $\la$ consists of only one part $n$ then
$Q_\la$  is $Q_n$ by above definition.

The elements $Q_\la$ constitute a basis of the vector space $\La\,$.
Furthermore, define a bilinear form $\langle\ ,\,\rangle_t$ 
on the vector space $\La$
by setting for any partitions $\la$ and $\mu$
\begin{equation}
\label{hlprod}
\langle\,p_\la,p_\mu\ts\rangle_t=k_\la\,\de_{\la\mu}\,
\prod_{i=1}^{\ell(\la)}\,\frac{1\,}{1-t^{\ts\la_i}}
\end{equation}
in the notation \eqref{schurprod}. It is obviously symmetric
and non-degenerate. By~[III.4.9]
\begin{equation}
\label{qq}
\langle\,Q_\la,Q_\mu\ts\rangle_t=b_\la(t)\,\de_{\la\mu}
\end{equation}
where
\begin{equation*}
\label{bla}
b_\la(t)\,=\,\prod_{i\ge1}\,\,\prod_{j=1}^{k_i}\,\,
(\ts 1-t^{\,j}\ts)\,.
\end{equation*}

Along with the symmetric function $Q_\la$ it is convenient to
use the symmetric function $P_\la$ which is a scalar multiple
of $Q_\la\ts$. By definition,
\begin{equation}
\label{qbp}
Q_\la=b_\la(t)\,P_\la 
\end{equation}
so that due to \eqref{qq}
$$
\langle\,P_\la,Q_\mu\ts\rangle_t=\de_{\la\mu}\,.
$$

%Note that at $t=0$ both $P_\la$ and $Q_\la$
%specialize to the \emph{Schur symmetric function}~$s_\la\ts$.
%In this special case the recursive formula \eqref{qla} belongs to
%Bernstein [Ex.\,I.5.29].

%------------------------------------------------------------------------------

\subsection{Macdonald functions}

Now let $\FF$ be the field $\QQ(q,t)$
with $q$ and $t$ parameters independent of each other. 
Generalizing \eqref{hlprod}
define a bilinear form $\langle\ ,\,\rangle_{q,t}$ on $\La$
by setting for any partitions $\la$ and $\mu$
\begin{equation}
\label{macprod}
\langle\,p_\la,p_\mu\ts\rangle_{q,t}=k_\la\,\de_{\la\mu}\,
\prod_{i=1}^{\ell(\la)}\,\frac{1-q^{\ts\la_i}}{1-t^{\ts\la_i}\,}
\end{equation}
in the notation of \eqref{schurprod}. This form is again symmetric
and non-degenerate. We will indicate by the superscript ${}^\ast$
the operator conjugation relative to the latter form. In particular, 
by \eqref{pperp} and \eqref{macprod}
for any $n\ge1$ we have 
\begin{equation*}
\label{past}
p_n^{\ts\ast}=\frac{1-q^n}{1-t^n}\,p^{\ts\perp}\,.
\end{equation*}
Hence by \eqref{qprod} we get
\begin{equation}
\label{qzexp}
Q^{\ast}(z)=1+Q_1^{\ast}\ts z+Q_2^{\ast}\ts z^2+\ldots
\,=\,\exp\,\Bigl(\,\,
\sum_{n\ge1}
\frac{1-q^n\!}{n}\,\ts p_n^{\ts\perp}\ts z^n
\ts\Bigr)\,.
\end{equation}

Using \eqref{qbp} when %the partition 
$\la$ consists of only one part $n$ we get
$P_n=Q_n\ts/(1-t)\,$.
Now consider the linear operator acting on the vector space $\La$ as the sum
\begin{equation}
\label{A1}
\sum_{n\ge1}\,
q^{\ts-n}\,Q_n\ts P_n^{\,\ast}=
\sum_{n\ge1}\,
q^{\ts-n}\,Q_n\ts Q_n^{\,\ast}\ts/\ts(1-t)\,.
\end{equation}
For future discussion
note that \eqref{A1} equals the coefficient at $1$ of the series~in~$z$
$$
(\ts Q(z)\,Q^*(\ts q^{-1}z^{-1})-1\ts)\ts/(1-t)\,.
$$ 
The operator \eqref{A1} is 
clearly self-conjugate relative to the bilinear form~\eqref{macprod}. 
By \cite[Eq.\,32]{AMOS} for any partition $\la$ 
the \emph{Macdonald symmetric function\/}  
$M_\la\in\La$ can be defined up to normalization
as an eigenvector of \eqref{A1} with the eigenvalue
\begin{equation}
\label{E1}
\sum_{i\ge1}\,(\ts q^{\ts-\la_i}-1\ts)\,t^{\,i-1}\,.
\end{equation}
For different partitions $\la$
the eigenvalues \eqref{E1} are pairwise distinct in $\QQ(q,t)\ts$.
It follows that the eigenvectors $M_\la$ with different $\la$ 
are pairwise orthogonal relative to \eqref{macprod}.
In the present article we will not be choosing any 
normalization of~$M_\la\ts$. We will use only the fact [VI.4.7] that 
the $M_\la$ form a basis of the vector space~$\La\,$.

%------------------------------------------------------------------------------

\subsection{Higher Hamiltonians}

In \cite{NS2} we introduced the following 
generalization of the operator \eqref{A1}. 
For any $k=0,1,2,\ldots$ consider the linear operator on $\La$
\begin{equation*}
\label{basic}
A^{\ts(k)}=\sum_{\ell(\la)=k}
q^{\,-\la_1-\la_2-\ts\ldots}\,Q_\la\ts P_\la^{\,\ast}.
\end{equation*}
Then $A^{\ts(0)}=1$ while $A^{\ts(1)}$ is the operator \eqref{A1}.
For any $k$ the operator $A^{\ts(k)}$ is 
obviously self-conjugate relative to \eqref{macprod}. 
Consider a series in a variable $u$
\begin{equation}
\label{au}
A(u)\,=\,\sum_{k\ge0}\,
{A^{\ts(k)}}/\ts{(\ts u\,;t^{\ts-1}\ts)_k}
\end{equation}
where as usual
$$
(\ts u\,;t^{\ts-1}\ts)_k\,=\,\prod_{j=0}^{k-1}\,(\ts1-u\,t^{\ts-j}\ts)\,.
$$
In \cite{NS2} we proved
\begin{equation}
\label{aigen}
A(u)\,M_\la\,=\,M_\la\,
\,\prod_{i\ge1}\,\,\frac{\,q^{\ts-\la_i}-u\,t^{\,1-i}}{1-u\,t^{\,1-i}}\,\,.
\end{equation}
If follows from \eqref{aigen} that
the operators $A^{\ts(1)},A^{\ts(2)},\,\ldots$ on $\La$  pairwise commute.
The eigenvalue \eqref{E1} of the operator
$A^{\ts(1)}$ can also be obtained from this equality.
 
The equality \eqref{aigen} can be derived 
from the results of \cite[Sec.\,3]{AK} 
which in turn are modifications
of those of \cite[Sec.\,9]{S}.
Our proof \cite[Sec.\,3]{NS2} was independent of all those results.
Let us now establish a relation between 
the works \cite{AK}~and~\cite{NS2}.

For $k\ge1$
denote by $S^{\ts(k)}$ the constant term of the formal series in $z_1\lcd z_k$
\begin{equation}
\label{sk}
Q(z_1)\ldots Q(z_k)\,
Q^*(q^{-1}z_1^{-1})\ldots Q^*(q^{-1}z_k^{-1})\,
F(z_1\lcd z_k)\,.
\end{equation}
Here the product \eqref{fuk} is regarded as a series in $z_1\lcd z_k$
using the expansion rule explained just after displaying it.
This $S^{\ts(k)}$ is a certain 
linear operator on the vector space $\La\,$.
It is convenient to set $S^{\ts(0)}=1\,$.
It turns out that $M_\la$ for each $\la$ is an eigevector
of the operators $S^{\ts(1)},S^{\ts(2)},\,\ldots$ like that of 
$A^{\ts(1)},A^{\ts(2)},\,\ldots\,\,$. 
This fact goes back to \cite{S}.
It also follows from \eqref{aigen} by the next proposition.

\begin{pro}
We have the relation
$$
(\ts u^{-1}\ts;t\,)_\infty\,A(u)\,=\,\sum_{k\ge0}\,\,
(\ts u\,t\ts)^{-k}\,S^{\ts(k)}/\ts{(\ts t^{\ts-1};t^{\ts-1}\ts)_k}
$$
where as usual
$$
(\ts u^{-1}\ts;t\,)_\infty
\,=\,\prod_{j=0}^\infty\,(\ts1-u^{-1}\ts t^{\,j}\,)\,.
$$
\end{pro}

\begin{proof}
For every partition $\la$ let us denote by $P_\la(z_1\lcd z_k)$ 
the specialization of the symmetric function 
$P_\la$ to 
$x_1=z_1\ts\lcd x_k=z_k$ and \text{$x_{k+1}=x_{k+2}=\ldots=0\,$.}
This is a homogeneous symmetric polynomial in the variables
$z_1\lcd z_k$ of degree $\la_1+\la_2+\ldots\,\,$. 
By using the first equality in 
\eqref{qprod} 
and then the expansion [III.4.4] %we get
$$
Q(z_1)\ldots Q(z_k)
\,=\,\prod_{i\ge1}\,\prod_{j=1}^k\ts
\frac{\,1-t\,x_i\,z_j\,}{1-x_i\ts z_j}
\,=\sum_{\ell(\la)\le k}Q_\la\,P_\la(z_1\lcd z_k)\,.
$$
It follows from the latter equality that 
$$
Q^*(q^{-1}z_1^{-1})\ldots Q^*(q^{-1}z_k^{-1})
\,=\sum_{\ell(\mu)\le k}
q^{\,-\mu_1-\mu_2-\ts\ldots}\,
P_\mu^{\,\ast}\,Q_\mu(z_1^{-1}\lcd z_k^{-1})\,.
$$
Hence 
$$
S^{\ts(k)}\,=\sum_{\ell(\la),\ell(\mu)\le k}
q^{\,-\mu_1-\mu_2-\ts\ldots}\,Q_\la\,P_\mu^{\,\ast}\,a_{\la\mu}(t)
$$
where $a_{\la\mu}(t)$ denotes
the constant term of the formal series in $z_1\lcd z_k$
$$
P_\la(z_1\lcd z_k)\,
Q_\mu(z_1^{-1}\lcd z_k^{-1})\,
F(z_1\lcd z_k)
\,.
$$
It is known that
$$
a_{\la\mu}(t)=\de_{\la\mu}\,
\frac{(\ts t\,;t\ts)_k}{(\ts t\,;t\ts)_{\ts k-\ell(\la)}}\,
$$
see \cite[App.\,B]{AK} for an elementary proof of this fact. 
Thus we obtain the relation
\begin{equation}
\label{stoa}
S^{\ts(k)}=\,
\sum_{i=0}^k\,
A^{\ts(i)}
\frac{(\ts t\,;t\ts)_k}{(\ts t\,;t\ts)_{\ts k-i}}\ .
\end{equation}

By substituting
the latter expression for $S^{\ts(k)}$
in our Proposition and by using the definition \eqref{au}
with the running index $k$ replaced by $i$ it remains to prove 
$$
\sum_{i\ge0}\,
{A^{\ts(i)}}\,
\frac{(\ts u^{-1}\ts;t\,)_\infty\ts}{(\ts u\,;t^{\ts-1}\ts)_{\ts i}}
\,=\,
\sum_{k\ge0}\,\,\sum_{i=0}^k\,A^{\ts(i)}\,
\frac{(\ts t\,;t\ts)_k}
{(\ts u\,t\ts)^{k}\,(\ts t^{\ts-1};t^{\ts-1}\ts)_k\,(\ts t\,;t\ts)_{\ts k-i}}
\ .
$$
By equating here the coefficients at $A^{\ts(i)}$ we have to prove
that for every $i\ge0$
$$
\frac{(\ts u^{-1}\ts;t\,)_\infty\ts}{(\ts u\,;t^{\ts-1}\ts)_{\ts i}}
\,=\,
\sum_{k\ge i}\ 
\frac{(\ts t\,;t\ts)_k}
{(\ts u\,t\ts)^{k}\,(\ts t^{\ts-1};t^{\ts-1}\ts)_k\,(\ts t\,;t\ts)_{\ts k-i}}
\ .
$$
But the last relation  follows 
by setting $v=u^{-1}\ts t^{\,i}$ and $j=k-i$
in the equality 
$$
(\ts v\,;t\,)_\infty
\,=\,
\sum_{j\ge0}\,
\frac{\ts(\ts-\ts v\ts)^{\ts j}\,t^{\,j(j-1)/2}}{(\ts t\,;t\ts)_j}\ .
\eqno{\qed}
$$
\end{proof}

Note that the relation \eqref{stoa} established above
is equivalent to \cite[Eq.\,3.3]{AK}. 
By using a variation of the
M\"obius inversion \cite[Lemma 5.1]{BGHT} we get from \eqref{stoa}
\begin{equation*}
\label{atos}
A^{\ts(k)}=\,
\sum_{i=0}^k\,
S^{\ts(i)}\,
\frac{\ts(\ts-\ts1\ts)^{\ts k-i}\,t^{\,(k-i)(k-i-1)/2}}
{(\ts t\,;t\ts)_i\,(\ts t\,;t\ts)_{\ts k-i}}\ .
\end{equation*} 
   
%We will use Proposition \ref{pro1} in the next section.

%==============================================================================

\section{Lax operator and Baker\ts-Akhiezer function}

%------------------------------------------------------------------------------

\subsection{Lax operator}

In this section we will construct yet another family 
%$I^{\ts(1)},I^{\ts(2)},\,\ldots$ 
of pairwise commuting operators 
on $\La$ with the Macdonald symmetric functions $M_\la$ 
being their eigenvectors. Let
$$
\La^\infty=z^{-1}\La\,[z^{-1}]
$$ 
be the ring of polynomials in $z^{-1}$ with the coefficients from
$\La$ but without the constant term. Introduce the linear operator $\,\U$ on 
the vector space $\La^\infty$ by setting
$$
\U:f(z)\mapsto [\ts Q(z)\ts f(z)\ts]_{\ts-}
$$
where the symbol $[\ \ ]_{\ts-}$ means
taking the only negative degree terms of 
the series. %~in~$z\,$.  

Let us extend the bilinear form \eqref{macprod} from $\La$ to $\La^\infty$ 
so that the subspaces 
\begin{equation}
\label{lasub}
z^{-1}\La\,,z^{-2}\La\,,\,\ldots\,\subset\,\La^\infty
\end{equation}
are orthogonal to each other, while each one carries the bilinear 
form determined by identifying that subspace with $\La\ts$.
For the operator $\Ua$ on $\La^\infty$ conjugate to $\U$ 
%relative to this form 
we then have 
$$
\Ua:f(z)\mapsto Q^\ast(z^{-1})\ts f(z)\ts.
$$
Moreover, using the decomposition of $\La^\infty$ into
the direct sum of subspaces \eqref{lasub}
the operators $\U$ and $\Ua$ are represented by infinite matrices
with operator entries
$$
\begin{pmatrix}
\,1&Q_1\!&Q_2&\cdots\  
\\[2pt]
\ 0\,&1&Q_1&\cdots 
\\[2pt]
\,0&0&1& \cdots
\\[-2pt]
\,\vdots&\vdots&\vdots&\ddots\,
\end{pmatrix}
\ \quad\text{and}\ \quad
\begin{pmatrix}
1&0&\,0&\cdots\ 
\\[2pt]
Q_1^*&1&\,0&\cdots 
\\[2pt]
Q_2^*&Q_1^*&\,1&\cdots
\\[-2pt]
\vdots&\vdots&\,\vdots&\ddots\,  
\end{pmatrix}
.
$$            

Further, let $\D_t$ be the linear operator on $\La^\infty$ defined by setting
$$
\D_t\ts:f(z)\mapsto f(z\,t^{-1})\,.
$$
Accordingly, let
$$
\D_q\ts:f(z)\mapsto f(z\,q^{-1})\,.
$$
The operators $\D_t$ and $\D_q$ are clearly self-conjugate relative 
to the bilinear form on $\La^\infty$ defined above.
They are represented by diagonal matrices with the entries
$t,t^2,\ldots$ and $q,q^2,\ldots$ respectively.
Our \emph{Lax operator\,} on $\La^\infty$ is the composition
\begin{equation*}
\label{laxoper}
\Lc=\Ua\D_t\,\U\,.
\end{equation*}

Furthermore introduce the operator $\Q:\La^\infty\to\La$ by setting
\begin{equation*}
\label{qc}
\Q:f(z)\mapsto [\ts Q(z)\ts f(z)\ts]_{\ts0}
\end{equation*}
where $[\ \,]_{\ts0}$ means
taking the constant term of a 
series~in~$z$. 
The conjugate operator $\Qa:\La\to\La^\infty$
is the application of $Q^\ast(z^{-1})-1$ to elements of $\La\,.$
The operators $\Q$ and $\Qa$ are represented by an infinite
row and a column with operator entries 
$$
(\,Q_1\,Q_2\,\ldots\,)
\ \quad\text{and}\ \quad
\begin{pmatrix}
Q_1^* 
\\[2pt] 
Q_2^* 
\\[-2pt]
\vdots
\end{pmatrix}
.
$$

Now put
\begin{equation}
\label{iu}
I(u)=\Q\,(u\,\D_q-\Lc\ts)^{-1}\Qa\,.
\end{equation}
It expands as a formal power series in $u^{-1}$ without constant term.
The coefficients of that series are self-conjugate operators on $\La$
by definition. 
These operators will form our new commuting family. 
The commutativity follows from the theorem below
which relates $I(u)$ to $A(u)\ts$.
The proof of the theorem shall be given later.

\begin{teo}
We have an equality of formal power series in $u^{-1}$
with operator coefficients acting on $\La$
\begin{equation*}
\label{ia}
\frac{u\,I(u)}{u-1}=1-\frac{A(u)}{A(u\,t^{-1})}
\,.\hspace{-20pt}
\end{equation*}
\end{teo}

By using this theorem, the eigenvalues of the operator coefficients of
the series $I(u)$ on the Macdonald symmetric functions $M_\la\in\La$
can be obtained from \eqref{aigen}.

%------------------------------------------------------------------------------

\subsection{Baker\ts-Akhiezer function}

Now introduce the formal power series in $u^{-1}$
\begin{equation}
\label{psi}
\Psi(u)=u\,(u\,\D_q-\Lc\ts)^{-1}\Qa A(u\,t^{-1})\,.
\end{equation}
The coefficients of this series are operators $\La\to\La^\infty$.
By the definitions \eqref{iu},\eqref{psi} %we have
\begin{equation}
\label{ipsi}
u\,I(u)\,A(u\,t^{-1})=\Q\ts\,\Psi(u)\,.
\end{equation}
We will call the series $\Psi(u)$ the \emph{Baker\ts-Akhiezer function\/}
for the Lax operator \eqref{laxoper}.
Our proof of Theorem 1 will be based on an expression for
$\Psi(u)$ given~next.

\begin{teo}
We have an equality of formal power series in $u^{-1}$
with operator coefficients mapping $\La$ to $\La^\infty$
\begin{equation*}
\label{eha}
\Psi(u)=E^\perp(-z^{-1})\,A(u\,t^{-1})\,H^\perp(z^{-1}q^{-1})-A(u\,t^{-1})
\,.
\end{equation*}
\end{teo}

We will prove Theorem 2 in the next subsection. 
Let us now derive Theorem~1 from it. Multiplying both sides of the relation 
in Theorem~1 by $(u-1)\, A(u\,t^{-1})$ on the right 
and then using \eqref{ipsi}                                                  
we get an equivalent relation to prove\ts:
\begin{equation}
\label{qpsi}
A(u\,t^{-1})+\Q\ts\,\Psi(u)=
u\, A(u\,t^{-1})-(u-1)\ts A(u)\,.
\end{equation}
But by using Theorem 2 along with definitions \eqref{ju},\eqref{qc}
we get the equalities
\begin{align*}
A(u\,t^{-1})+\Q\ts\,\Psi(u)
&=[\,Q(z)\,E^\perp(-z^{-1})\,A(u\,t^{-1})\,H^\perp(z^{-1}q^{-1})\ts]_{\ts0}
\\
&=[\,J(z)\,A(u\,t^{-1})\,H^\perp(z^{-1}q^{-1})\ts]_{\ts0}\ts.
\end{align*}
By our Proposition, the right hand of these equalities
can be rewritten~as~the~sum
\begin{equation}
\label{jsh}
\sum_{k\ge0}\,\,
u^{-k}\,[\,J(z)\,S^{\ts(k)}\,H^\perp(z^{-1}q^{-1})\ts]_{\ts0}
\ts/\ts{(\ts t^{\ts-1};t^{\ts-1}\ts)_k}
\end{equation}
divided by $(\ts u^{-1}\ts t\,;t\,)_\infty\,$.
Note that by using %the identities 
\eqref{euexp},\eqref{huexp}
and then %the identity 
\eqref{qzexp} we have
$$
E^\perp(-z^{-1})\,H^\perp(z^{-1}q^{-1})
\,=\,\exp\,\Bigl(\,\,
\sum_{n\ge1}
\frac{1-q^n\!}{n}\,\ts p_n^{\ts\perp}\ts z^{-n}q^{-n}
\ts\Bigr)
=\,
Q^\ast(z^{-1}q^{-1})
\,.
$$
Therefore by recalling 
the definition of the operator $S^{\ts(k)}$ on $\La$
and then by using the comultiplicativity [Ex.\,I.5.25] of
$E^\perp(-z^{-1})\ts$, the factor
$$
[\,J(z)\,S^{\ts(k)}\,H^\perp(z^{-1}q^{-1})\ts]_{\ts0}
$$
in the summand of \eqref{jsh} is equal to
the constant term of the series in $z_1\lcd z_k$
\begin{gather*}
[\,J(z)\ts(\ts Q(z_1)\ldots Q(z_k))\,
Q^\ast(z^{-1}q^{-1})\,
Q^*(q^{-1}z_1^{-1})\ldots Q^*(q^{-1}z_k^{-1})\ts]_{\ts0}\,
F(z_1\lcd z_k)\,.
\end{gather*}
By \eqref{qkz} and \eqref{fuk}
this constant term is exactly $S^{\ts(k+1)}$. 
So the sum \eqref{jsh}~equals
$$
\sum_{k\ge0}\,\,
u^{-k}\,S^{\ts(k+1)}
/\ts{(\ts t^{\ts-1};t^{\ts-1}\ts)_k}
\,\,=\,\,
\sum_{k\ge0}\,\,
u^{\ts1-k}\,(1-t^{-k})\,S^{\ts(k)}
/\ts{(\ts t^{\ts-1};t^{\ts-1}\ts)_k}
$$
where we first replaced $k+1$ by $k$
and then formally included the zero summand corresponding to $k=0\ts$.
Using our Proposition, the last displayed sum %over $k\ge0$
equals %the difference
$$
(\ts u^{-1}\ts t\,;t\,)_\infty\,u\,A(u\,t^{-1})
-(\ts u^{-1}\ts;t\,)_\infty\,u\,A(u)\,.
$$
Dividing this difference by $(\ts u^{-1}\ts t\,;t\,)_\infty$
we get the right hand side of the relation \eqref{qpsi}.
This we have derived Theorem 1 from Theorem 2.
Let us prove the latter.

%------------------------------------------------------------------------------

\subsection{Proof of Theorem 2}
Due to the definition \eqref{psi} we have to prove the relation  
$$
u\,(u\,\D_q-\Lc\ts)^{-1}\Qa A(u\,t^{-1})
=
E^\perp(-z^{-1})\,A(u\,t^{-1})\,H^\perp(z^{-1}q^{-1})-A(u\,t^{-1})\,.
$$
Let us multiply both sides of this relation by $u\,\D_q-\Lc$ on the left.  
In this way
we get an equivalent relation to prove\ts:
$$
u \,\Qa A(u\,t^{-1})=(u\,\D_q-\Lc\ts)\,
(\ts E^\perp(-z^{-1})\,A(u\,t^{-1})\,H^\perp(z^{-1}q^{-1})-A(u\,t^{-1}))\,.
$$
By the definitions of the operators  $\Qa$ and $\Lc$ the latter relation 
can be written~as 
\begin{gather*}
u\,(\ts Q^\ast(z^{-1})-1\ts)\,A(u\,t^{-1})=
\\
(u\,\D_q-\Ua\D_t\,\U\ts)\,
(\ts E^\perp(-z^{-1})\,A(u\,t^{-1})\,H^\perp(z^{-1}q^{-1})-A(u\,t^{-1}))\,.
\end{gather*}
Using the  definitions of the operators $\D_q\ts,\D_t$ and $\U\ts,\ts\Ua$ 
it can be rewritten as
\begin{gather*}
u\,(\ts Q^\ast(z^{-1})-1\ts)\,A(u\,t^{-1})=
u\,(\ts E^\perp(-z^{-1}q)\,A(u\,t^{-1})\,H^\perp(z^{-1})-A(u\,t^{-1}))
\\
-\,Q^\ast(z^{-1})\,[\,
Q(z\,t^{-1})\,E^\perp(-z^{-1}t)\,A(u\,t^{-1})\,H^\perp(z^{-1}t\,q^{-1})
-A(u\,t^{-1}))\ts]_{\ts-}\,.
\end{gather*}
Here both sides are series in $u$ with operator coefficients that map
$\La$ to $\La^\infty$. We can also regard
both sides as series in $u$ and $z$ with operator coefficients 
mapping $\La$ to $\La\ts$. This allows us to perform obvious cancellations
hence getting to prove
\begin{gather*}
u\,Q^\ast(z^{-1})\,A(u\,t^{-1})=
u\,E^\perp(-z^{-1}q)\,A(u\,t^{-1})\,H^\perp(z^{-1})
\\
-\,Q^\ast(z^{-1})\,[\,
Q(z\,t^{-1})\,E^\perp(-z^{-1}t)\,A(u\,t^{-1})\,
H^\perp(z^{-1}t\,q^{-1})\ts]_{\ts-}\,.
\end{gather*}

Using the definition \eqref{ju} the last displayed relation can be also 
written as 
\begin{gather}
\nonumber
u\,Q^\ast(z^{-1})\,A(u\,t^{-1})=
u\,E^\perp(-z^{-1}q)\,A(u\,t^{-1})\,H^\perp(z^{-1})
\\
\label{reduced}
-\,Q^\ast(z^{-1})\,[\,
J(z\,t^{-1})\,A(u\,t^{-1})\,H^\perp(z^{-1}t\,q^{-1})\ts]_{\ts-}\,.
\end{gather}
Here $A(u\,t^{-1})$ is a formal power series in $u^{-1}$ with leading term $1$
by \eqref{au}.~We also know that
\begin{equation}
\label{ehqz}
E^\perp(-z^{-1}q)\,H^\perp(z^{-1})=Q^\ast(z^{-1})\,,
\end{equation}
see the previous subsection. Hence the coefficents at $u$ of both sides of the
relation \eqref{reduced} coincide. Let us now multiply both sides of the 
relation \eqref{reduced} by $(\ts u^{-1}\ts;t\,)_\infty$ and 
take the coefficients at $u^{\ts k-1}$ for any $k\ge1\ts$.
Due to our Proposition we~obtain
\begin{gather*}
Q^\ast(z^{-1})\,S^{\ts(k)}/\ts{(\ts t^{\ts-1};t^{\ts-1}\ts)_k}=
E^\perp(-z^{-1}q)\,
S^{\ts(k)}H^\perp(z^{-1})/\ts{(\ts t^{\ts-1};t^{\ts-1}\ts)_k}
\\
-\,Q^\ast(z^{-1})\,[\,
J(z\,t^{-1})\,S^{\ts(k-1)}H^\perp(z^{-1}t\,q^{-1})\ts]_{\ts-}\ts
/\ts{(\ts t^{\ts-1};t^{\ts-1}\ts)_{k-1}}\,.
\end{gather*}
Next multiply both sides by ${(\ts t^{\ts-1};t^{\ts-1}\ts)_k}$ and divide by
$Q^\ast(z^{-1})$ on the left.~We~get
\begin{gather}
\notag
S^{\ts(k)}=Q^\ast(z^{-1})^{-1}
E^\perp(-z^{-1}q)\,S^{\ts(k)}H^\perp(z^{-1})
\\
-\,(\ts1-t^{-k}\ts)\,[\,
J(z\,t^{-1})\,S^{\ts(k-1)}H^\perp(z^{-1}t\,q^{-1})\ts]_{\ts-}\,.
\label{divided}
\end{gather}

The operator $S^{\ts(k)}$ at the left hand side of the relation 
\eqref{divided} is
the constant term of the series \eqref{sk} in $z_1\lcd z_k\,$.
Using this definition along with \eqref{hq},\eqref{ehqz} and the 
comultiplicativity of 
$H^\perp(z^{-1})\ts$, the first summand of the right hand side of 
\eqref{divided}
is equal to the sum of those terms of the formal series in $z,z_1\lcd z_k$
$$
Q(z_1)\ldots Q(z_k)\,
Q^*(q^{-1}z_1^{-1})\ldots Q^*(q^{-1}z_k^{-1})\,
F(z_1\lcd z_k)\,
\prod_{j=1}^k\,\frac{z-z_j}{z-t\,z_j}\ .
$$
that are  free of $z_1\lcd z_k\,$.
To present in a similar way the expression 
displayed in the second line of \eqref{divided} we will use
the following simple lemma.
Let $G(z)$ be a formal series in $z$ with coefficients in any algebra over 
%the field
$\QQ(t)$. Let $[\ts G(z)\ts ]_{\ts0}$ be the constant term of this series.
Let $[\ts G(z)\ts ]_{\ts-}$ be the sum of the negative degree~terms.

\begin{lem}
The sum
$$
[\ts G(z)\ts ]_{\ts0}+(1-t^{-1})\,[\,G(\ts z\,t^{-1}\ts)\ts ]_{\ts-}\!
$$
is equal to the sum of those terms of the series in $z\ts,w$
$$
G(w)\,\frac{z-w}{z-t\,w}
$$
that are free of $w\ts$. Here the fraction should be expanded
as a power series~in~$w/z\,$.
\end{lem}

Verifying this lemma is straighforward. Let us now set
$$
G(z)=J(z)\,S^{\ts(k-1)}H^\perp(z^{-1}q^{-1})\,.
$$
By the definition of $S^{\ts(k-1)}$ this $G(z)$ is the sum of
those terms of the~series
$$
J(z)\,
Q(z_1)\ldots Q(z_{k-1})\,
Q^*(q^{-1}z_1^{-1})\ldots Q^*(q^{-1}z_{k-1}^{-1})\,
H^\perp(z^{-1}q^{-1})
F(z_1\lcd z_{k-1})\,
$$
in $z,z_1\lcd z_{k-1}$ that are free of $z_1\lcd z_{k-1}\,$.
The last displayed series equals
\begin{gather*}
Q(z)\,
Q(z_1)\ldots Q(z_{k-1})\,
Q^*(q^{-1}z^{-1})\,
Q^*(q^{-1}z_1^{-1})\ldots Q^*(q^{-1}z_{k-1}^{-1})\ \times
\\
F(z,z_1\lcd z_{k-1})
\end{gather*}
by \eqref{qkz} and \eqref{fuk}. Note that we  have used a similar argument
in the previous subsection. Denote
$$
c_k(t)=(1-t^{k})/(1-t)\,.
$$
Applying the lemma, the expression 
displayed in the second line of \eqref{divided} is equal the constant term
of the series in $z,z_1\lcd z_{k-1}$
\begin{gather}
\notag
Q(z)\,
Q(z_1)\ldots Q(z_{k-1})\,
Q^*(q^{-1}z^{-1})\,
Q^*(q^{-1}z_1^{-1})\ldots Q^*(q^{-1}z_{k-1}^{-1})\ \times
\\
\label{z}
c_k(t^{-1})\,
F(z,z_1\lcd z_{k-1})\,\,
\end{gather}
minus those terms
of the series in $z,w,z_1\lcd z_{k-1}$
\begin{gather}
\notag
Q(w)\,
Q(z_1)\ldots Q(z_{k-1})\,
Q^*(q^{-1}w^{-1})\,
Q^*(q^{-1}z_1^{-1})\ldots Q^*(q^{-1}z_{k-1}^{-1})\ \times
\\
\label{w}
c_k(t^{-1})\,
F(w,z_1\lcd z_{k-1})\,
\frac{z-w}{z-t\,w}\,
\end{gather}
that are free of $w,z_1\lcd z_{k-1}\,$. As we are taking the constant
term,
the variables $z,z_1\lcd z_{k-1}$ in \eqref{z} can be replaced by
$z_1\lcd z_{k}$ respectively.
The variables $w,z_1\lcd z_{k-1}$ in \eqref{w} can be replaced by
$z_1\lcd z_{k}$ as well.

Recall that the coefficients $Q_1,Q_2,\ldots$ of the series $Q(z)$ are free
generators of the algebra 
$\La\,$, while the operator product
$$
Q(z_1)\ldots Q(z_k)\,
Q^*(q^{-1}z_1^{-1})\ldots Q^*(q^{-1}z_k^{-1})\,
$$
is symmetric in $z_1\lcd z_{k}\,$. By the above presentation of the terms of
\eqref{divided}, that relation is equivalent to the equality between the 
symmetrization
of $F(z_1\lcd z_k)$ and that of 
\vspace{-8pt}
\begin{gather}
\notag
F(z_1\lcd z_k)\,
\prod_{j=1}^k\,\frac{z-z_j}{z-t\,z_j}\ +
\\
\label{s}
c_k(t^{-1})\,
F(z_1\lcd z_k)\,
-\,c_k(t^{-1})\,
F(z_1\lcd z_k)\,
\frac{z-z_1}{z-t\,z_1}\ .
\end{gather}
Here symmetrizing means taking 
the sum over all $k\ts!$ permutations of $z_1\lcd z_k\ts$.

Let us prove the latter equality. 
At $z=\infty$ the sum \eqref{s} equals $F(z_1\lcd z_k)$ even before 
symmetrization.
We may assume that $z_1\lcd z_k$ are pairwise distinct.
Then it suffices to check that 
that the symmetrization of \eqref{s} has no poles at 
$z=t\ts z_1\lcd t\ts z_k\,$.
By the symmetry in $z_1\lcd z_k$ taking only $z=t\ts z_1$ will suffice. 

The product over $j=1\lcd k$ showing in the first line of \eqref{s}
is symmetric in $z_1\lcd z_k\ts$. At $z=t\ts z_1$ it has a simple pole.
By [III.1.4] the symmetrization of $F(z_1\lcd z_k)$ is
$$
c_1(t)\ldots c_k(t)\,\prod_{\substack{1\le i,j\le k\\i\neq j}}\,
\frac{z_i-z_j}{\,z_i-t\,z_j}\ .
$$
Hence the residue at $z=t\ts z_1$
of the symmetrization of
the whole expression in the first line of \eqref{s} is
\begin{equation}
\label{canone}
c_1(t)\ldots c_k(t)\,(t-1)\,z_1\,
\prod_{\substack{1\le i,j\le k\\i\neq j}}\,\frac{z_i-z_j}{\,z_i-t\,z_j}
\ \cdot\ 
\prod_{j=2}^k\,\frac{t\,z_1-z_j}{t\,(z_1-z_j)}\ .
\end{equation}
When symmetrizing the negative term of the difference
displayed in the second line of \eqref{s}, we get a pole
at $z=t\ts z_1$ only from the permutations of $z_1\lcd z_k$ 
preserving $z_1\ts$. Applying [III.1.4] 
once again but to the $k-1$ variables $z_2\lcd z_k$
instead of  $z_1\lcd z_k$
the residue  at $z=t\ts z_1$ 
of the symmetrization of the difference displayed in
the second line of \eqref{s} is
\begin{equation}
\label{cantwo}
-\,c_k(t^{-1})\,
c_1(t)\ldots c_{k-1}(t)\,(t-1)\,z_1\,
\prod_{\substack{2\le i,j\le k\\i\neq j}}\,\frac{z_i-z_j}{\,z_i-t\,z_j}
\ \cdot\ 
\prod_{j=2}^k\,\frac{z_1-z_j}{\,z_1-t\,z_j}\ .
\end{equation}
To verify that the sum of two products \eqref{canone} and \eqref{cantwo} 
is zero, we can~cancel in both of them the product of the common factors
$$
c_1(t)\ldots c_{k-1}(t)\,(t-1)\,z_1\,
\prod_{\substack{2\le i,j\le k\\i\neq j}}\,\frac{z_i-z_j}{\,z_i-t\,z_j}
$$
and then use the relation $c_k(t)\,t^{\ts1-k}=c_k(t^{-1})\,$.
This verification completes our proof of Theorem~2.

%==============================================================================

\section*{\normalsize\bf Acknowledgements}

The first named author has been supported
by EPSRC grant EP/I\,014071, and also
acknowledges support of IH\'ES
where he stayed in October\ts-November 2014.

%=============================================================================

%=============================================================================

\end{document}